\documentstyle[12pt,epsf]{article}

\newcommand{\ket}[1]{| #1 \rangle}

\begin{document}

\begin{center}
{\large\bf Large Amplitude Collective Motion
in Nuclei and Metallic Clusters}\footnote{
Talk presented at Conference on
``Atomic Nuclei and Metallic Clusters: Finite Many-Fermion Systems''
(September 1 -- 5, 1997, Prague, Czech Republic)}\\
{\large --- Applicability of adiabatic theory for a pairing model ---}

\medskip
{Takashi~Nakatsukasa and Niels~R.~Walet }\\

{\it Department of Physics, UMIST, P.O.Box 88,
Manchester M60 1QD, UK}

\end{center}

\begin{abstract}
A model Hamiltonian describing a two-level system with a crossing
plus a pairing force
is investigated using technique of large-amplitude collective motion.
The collective path,
which is determined by the decoupling conditions,
is found to be almost identical
to the one in the Born-Oppenheimer approximation
for the case of a strong pairing force.
For the weak pairing case,
the obtained path describes a diabatic dynamics of the system.
\end{abstract}
\noindent
PACS number(s): 21.60.-n, 36.40.Sx

\bigskip

Nuclei and metallic clusters are finite fermionic many-body systems
which support many kinds of collective motion.
While properties of high-frequency vibrations (giant resonances)
may be well reproduced by the random-phase approximation (RPA),
some low-frequency vibrations exhibit an anharmonic nature
that the RPA cannot describe.
Shape coexistence and fission phenomena in nuclei and clusters
also have such a large amplitude nature.
In order to investigate
these kinds of large amplitude collective motion,
one would like to reduce the number of degrees of freedom to a few
judiciously chosen collective coordinates.

In order to determine the collective coordinates self-consistently,
we use the theory of adiabatic large amplitude collective motion
(ALACM) [1].
This provides a method to find decoupled motion
which is confined to a few-dimensional submanifold
of the configuration space.
In this theory,  a classical Hamiltonian is expanded with respect
to momenta up to the second order.
Since we are neglecting higher momentum terms,
this theory is based on the adiabatic (slow velocity) assumption.
However, since a nucleus is made of particles
which have almost the same mass,
the adiabatic assumption is never trivial, as in molecular physics.
In this paper, we will show that the ALACM can well
describe a system, even when the conventional adiabatic theory
(Born-Oppenheimer approximation) completely fails.

In nuclear phenomena,
it is well-known that the pairing influences all kinds of
low-frequency collective motion.
A well-known example is the moment of inertia for rotational nuclei,
which is always smaller than the rigid-body value at low spin.
This is supposed to be due to the paring correlations.
For nuclear spontaneous fission, 
the observed lifetimes of odd-$A$ nuclei are some orders of magnitude
larger than those of even-even nuclei.
This property has been understood in terms of a concept called
``specialization energy'' which is associated with the pairing.
The pairing force is considered to be responsible for
change in the particle configuration at level crossings.
Since the pairing interaction cannot act on the last odd particle,
this particle tends to stay in the same level even after the
crossing, which may lead to a higher potential barrier.

According to these considerations, the pairing should play a 
key role in understanding large amplitude collective motion,
especially when level crossings are involved
as the shape change is taking place.
In this paper, we apply the theory of ALACM
to a simple model Hamiltonian of a two crossing levels
with a pairing interaction.
The model Hamiltonian,
which has been introduced in Ref. [2],
is given by
\begin{eqnarray}
\label{H}
H  &=& \frac{1}{2} \left( p^2 + q^2 \right) + H_{\rm val} \ ,\\
\label{H_val}
H_{\rm val}  &=& ( \chi q - \epsilon_0 )
  \left( \hat{N}_1 - \hat{N}_2 \right)
        - G \hat{P}^\dagger \hat{P} \ ,
\end{eqnarray}
where 
$\hat{N}_\mu \equiv
c_\mu^\dagger c_\mu + c_{\bar \mu}^\dagger c_{\bar \mu}$
and
$\hat{P}^\dagger \equiv
\sum_{\mu=1,2} c_\mu^\dagger c_{\bar \mu}^\dagger$.
Here $\bar\mu$ denotes the time-reversed state of $\mu$
and we consider a valence space of two particles.
The valence particles are coupled to the core and
a crossing of single-particle levels occurs at
$q=\epsilon_0/\chi$.

In order to apply the ALACM theory,
we need to recast the quantum dynamics in
a classical Hamiltonian form [3].
Here we use the time-dependent mean field (BCS) theory
to derive a classical Hamiltonian corresponding to
$H_{\rm val}$.
In this case, if we consider the particle-number conservation,
the system can be described by two coordinates $(q,\xi)$ and their
conjugate momenta $(p,\pi)$.
Here the coordinate $q$ describes
the harmonic oscillator core in Eq.(\ref{H})
and $\xi$ describes the distribution of particles among two levels.

\begin{figure}
\epsfxsize=\textwidth
\epsffile{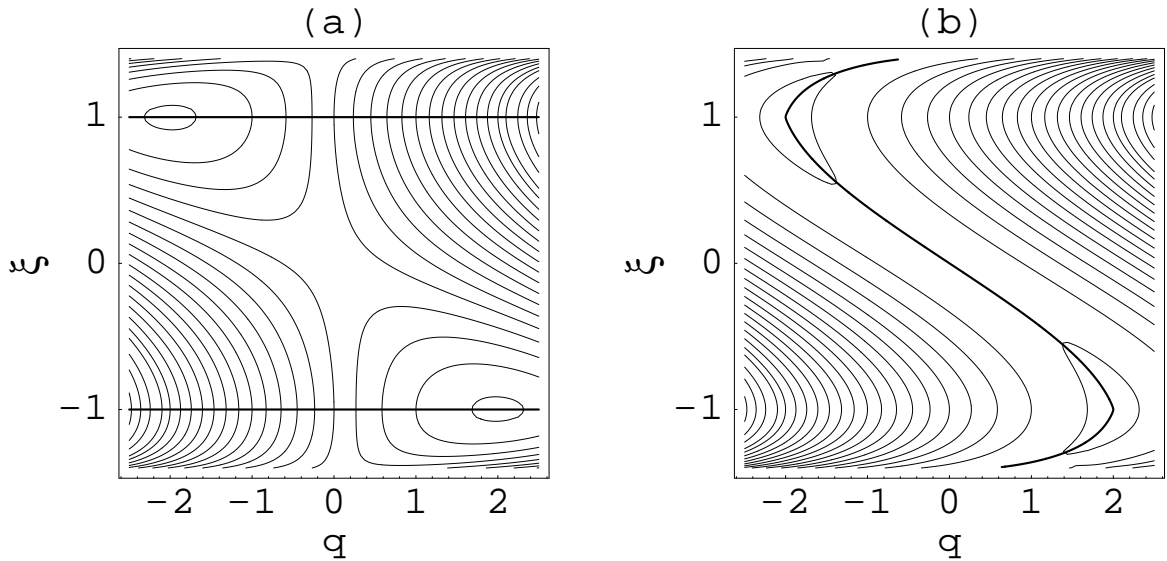}
\vspace{-0.5cm}
\centerline{
\begin{minipage}[t]{0.9\textwidth}
\caption{\small
Contour plots of energy surface for the potential $V(q,\xi)$.
The parameters $\epsilon_0=0$ and $\chi=1$ are used.
The left figure (a) shows the energy surface for $G=0.2$ and
the right (b) for  $G=1.9$.
Contour lines are displayed for $\Delta V = 0.5$.
Thick lines indicate the collective paths obtained by the ALACM
theory.
}
\end{minipage}
}
\end{figure}

Taking $\epsilon_0=0$ and $\chi=1$,
the potential $V(q,\xi)$ has two local minima at
$(q,\xi)=(-2,1)$ and $(2,-1)$ which correspond to
valence-particle configurations
$\ket{1\bar 1} \equiv c_1^\dagger c_{\bar 1}^\dagger \ket{0}$
and $\ket{2\bar 2}$, respectively.
A collective path is now obtained by solving
the local harmonic equations (LHE)
given in Ref. [1].
Since this system has only two degrees of freedom,
we can easily visualize the path on the 2-dimensional surface.
In Fig.~1, the potential energy surface and
the obtained collective path are shown
for two strengths of the paring force $G=0.2$ and $G=1.9$.
In this case ($\epsilon_0=0$),
the potential landscape is symmetric
with respect to a $180^\circ$ rotation around the origin,
$V(q,\xi) = V(-q,-\xi)$.
For a weak pairing force (a),
each local minimum has an independent collective path
which represents a harmonic oscillator with a fixed valence
configuration (both particles in level 1 or in level 2).
These represent diabatic solutions which do not
mix in this approximation.
On the other hand, for a strong pairing force (b),
we get a collective path which changes the particle configuration
and connects the two local minima.

In order to study the case of an asymmetric potential landscape,
let us take
\begin{equation}
\label{asym_para}
\epsilon=0.4, \quad \chi=0.7, \quad\mbox{and}\quad G=0.1 \ .
\end{equation}
First we adopt the conventional Born-Oppenheimer (BO) theory
by diagonalizing the valence Hamiltonian $H_{\rm val}$
at each value of $q$.
The resulting BO potential and wave functions for the ground state
and the second excited state are shown in Fig.~2.
If one exactly solves this problem quantum mechanically,
the valence-particle configuration of
the second excited state consists of
$\ket{2\bar 2}$ (99\%) and $\ket{1\bar 1}$ (1\%).
Therefore, the wave function should be well localized around
a shallow potential minimum at $q=2\chi=1.4$.
On the other hand,
the BO wave function spreads over a much wider range of $q$,
which apparently indicates a failure of the BO approximation
[2].
If we solve the LHE, however, the collective paths again
turn out to be two straight lines, $\xi=-1$ and $\xi=1$,
similar to Fig.~1 (a).
These paths lead to completely diabatic dynamics.
This result qualitatively agrees with the exact solution.

\begin{figure}
\begin{minipage}[t]{0.5\textwidth}
\epsfysize=10cm
\epsffile{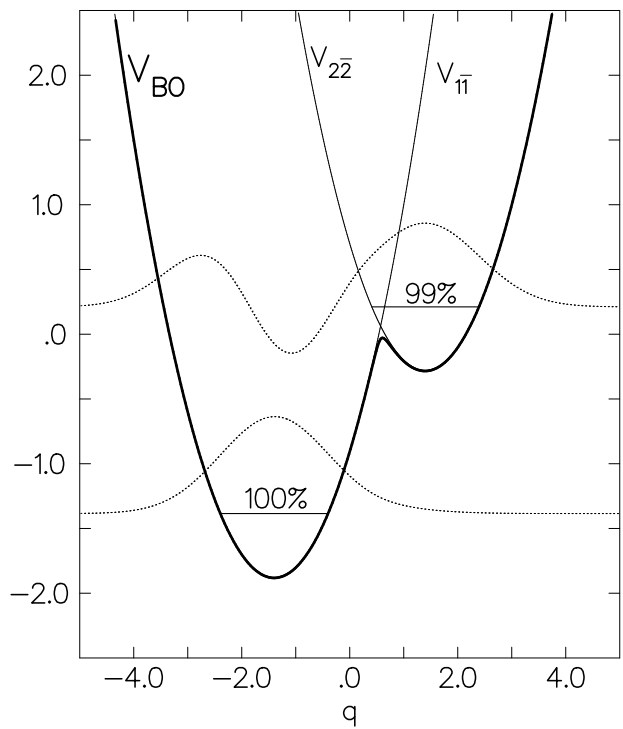}
\end{minipage}
\begin{minipage}[b]{0.47\textwidth}
\caption{\small
Thick line shows Born-Oppenheimer potential $V_{\rm BO}(q)$
for the parameters of Eq. (\ref{asym_para}).
Thin lines give the diabatic potential curves,
$V_{1\bar 1}$ and $V_{2\bar 2}$, with fixed valence configurations
$\ket{1\bar 1}$ and $\ket{2\bar 2}$, respectively.
The dotted lines indicate wave functions obtained
in the BO approximation
for the ground state and the second excited state.
The number on each level indicates the percentage of the main
configuration in the exact wave function.
}
\end{minipage}
\end{figure}

In conclusion,
we have found that the system automatically selects
different type of collective paths according to the strength of
pairing force.
In the case of weak pairing force,
the motion along a coordinate $q$ is not necessarily slower than
the motion along $\xi$,
which results in breakdown of the BO approximation.
In the ALACM theory, two coordinates, $q$ and $\xi$,
are treated equivalently
and a decoupled path is determined in the two dimensional space.
Then, the {\it adiabatic} theory may account for
the {\it diabatic} dynamics.
Due to limitation of space, we have not been able to give
many technical details.
These will appear in a future publication [4].
Applications to fissioning clusters are also envisaged.

This work is supported by EPSRC (UK).


\bigskip

\noindent
{\large\bf References}

\medskip
\noindent
[1] A.~Klein, N.R.~Walet and G.~Do~Dang, Ann. Phys. {\bf 208} (1991) 90.
\newline
[2] T.~Fukui, M.~Matsuo, K.~Matsuyanagi,
Prog. Theor. Phys. {\bf 85} (1991) 281.\newline
[3] J.-P.~Blaizot and G.~Ripka, {\it Quantum Theory of Finite Systems},
(The MIT Press, Cambridge, Massachusetts, 1986), p.298.\newline
[4] T.~Nakatsukasa and N.R.~Walet, in preparation.

\end{document}